\documentclass[useAMS,usenatbib]{mn2e}

\usepackage{graphicx}
\usepackage{amsmath}
\usepackage{amssymb}
\usepackage[usenames,dvipsnames]{color}
\usepackage[T1]{fontenc}
\usepackage[utf8]{inputenc}

\usepackage{hyperref}
\hypersetup{colorlinks=true,linkcolor=blue,citecolor=blue,filecolor=blue,urlcolor=blue}

\usepackage{aas_macros}

\newcommand\msun{M_\odot} 

\newcommand\kms{\rm km\,s^{-1}}
\newcommand\pc{{\rm pc}}

\newcommand\gyr{{\rm Gyr}}

\newcommand\rhini{r_{\rm h,ini}} 
\newcommand\fbin{f_{\rm bin}} 
\newcommand\Nbin{N_{\rm bin}} 
\newcommand\Nstar{N_{\rm star}} 
\newcommand\mini{M_{\rm ini}} 
\newcommand\nbseven{{\tt NBODY7}}
\newcommand\nbsix{{\tt NBODY6}}
\newcommand\bse{{\tt BSE}}
\newcommand\sse{{\tt SSE}}

\title[NSs and MSPs in star clusters]{Neutron stars and millisecond pulsars in star clusters: implications for the diffuse $\gamma$-radiation from the Galactic Centre}
\author[G. Fragione, V. Pavl{\'i}k \& S. Banerjee]{Giacomo Fragione$^{1}$\thanks{E-mail: giacomo.fragione@mail.huji.ac.il},
V{\'a}clav Pavl{\'i}k$^{2,3}$
\& Sambaran Banerjee$^{4,5}$\\
$^{1}$Racah Institute for Physics, The Hebrew University, Jerusalem 91904, Israel\\
$^{2}$Astronomical Institute of Charles University, V Hole{\v s}ovi{\v c}k{\'a}ch 2, 180 00 Prague 8, Czech Republic\\
$^{3}$Observatory and Planetarium of Prague, Kr{\'a}lovsk{\'a} obora 233, 170 21 Prague 7, Czech Republic\\
$^{4}$Argelander-Institut f{\"u}r Astronomie (AIfA), Auf dem H{\"u}gel 71, 53121 Bonn, Germany\\
$^{5}$Helmholtz-Instituts f{\"u}r Strahlen- und Kernphysik (HISKP), Nussallee 14-16, 53115 Bonn, Germany
}

\begin{document}

\maketitle

\begin{abstract}
Globular clusters (GCs) are the ideal environment for the formation of neutron stars (NSs) and millisecond pulsars (MSPs). NSs origin and evolution provide a useful information on stellar dynamics and evolution in star clusters. NSs are among the most interesting astrophysical objects, being precursors of several high-energy phenomena such as gravitational waves and gamma-ray bursts. Due to a large velocity kick that they receive at birth, most of the NSs escape the local field, affecting the evolution and dynamics of their parent cluster. In this paper, we study the origin and dynamical evolution of NSs within GCs with different initial masses, metallicities and primordial binary fractions. We find that the radial profile of NSs is shaped by the BH content of the cluster, which partially quenches the NS segregation until most of the BHs are ejected from the system. Independently on the cluster mass and initial configuration, the NSs map the average stellar population, as their average radial distance is $\approx 60-80\,\%$ of the cluster half-mass radius. Finally, by assuming a recycling fraction of $f_\mathrm{rec}=0.1$ and an average MSP gamma-ray emission of $L_\gamma=2\times 10^{33}$\,erg\,s$^{-1}$, we show that the typical gamma-ray emission from our GCs agrees with observations and supports the MSP origin of the gamma-ray excess signal observed by the \textit{Fermi-LAT} telescope in the Galactic Centre.
\end{abstract}


\section{Introduction}
\label{sect:intro}

Globular clusters (GCs) are known to be the breeding grounds for the formation of neutron stars (NSs) and millisecond pulsars (MSPs). GCs host a large number of NSs that are the final product of the lifetime of stars with masses between $\approx 9-20\,\msun$ \citep{hur00,hur02}. NSs are observed in GCs mainly through their participation in interacting binary systems, either as low-mass \hbox{X-ray} binaries (LMXBs) or MSPs, with typical luminosities of \hbox{$\approx 10^{31}$ -- $10^{38}$\,erg\,s$^{-1}$} in X- and gamma-rays \citep{iva08}. As a consequence of high densities in GCs, it has been observed that the abundance of these objects (per unit mass) is $\approx 100$ times higher than in the Galaxy as a whole \citep{kat75}.

NS origin and evolution provide us with useful information and constraints for our understanding of both stellar dynamics and evolution in star clusters. Even though hundreds of NSs are formed in a star cluster, most of them receive a velocity kick at birth whose magnitude is typically much larger than what is required to escape from a GC \citep{dav98,han97,Hobbs_2005}. Therefore, only a small fraction of NSs is finally retained in the cluster. \citet*{tre10} presented \hbox{$N$-body} simulations of star clusters with different primordial binary populations and mass profiles, and showed that the retention fractions of NSs and black holes (BHs) play a role in determining the final properties of the cluster, e.g.\ the size of its core. 

MSPs are believed to be ``recycled'' pulsars. This terminology comes from the most common and accepted pathway to form such objects, i.e.\ the spin up of a NS that accretes mass from a companion star in a binary system \citep{alp82,phi94}.
The accretion phase lasts few hundred Myrs, during which the system radiates and can be observed in X-ray \citep{tau12}. The formation of a MSP still remains a complex problem; a lot of physics takes place in speeding up the NS, which has not yet been understood completely \citep{tau11,tau12}.

New attention to NSs and MSPs comes from the recent high-quality data (in the energy range from about $20$\,MeV to over $300$\,GeV) from the Large Area Telescope instrument on board the \textit{Fermi} Gamma-Ray Space Telescope (Fermi-LAT). Fermi data proved the presence of an extended gamma-ray excess around the Galactic Centre, peaking at $\approx 2$\,GeV, with an approximately spherical density profile $\propto r^{-2.4}$ \citep{aba14,cal15}. The excess extends up to $\sim 20^\circ$, but the evidence for an extension beyond $\sim 10^\circ$ is quite weak, given the foreground subtraction systematics. There is no clear understanding of the origin of such a gamma-ray excess, with dark matter annihilation, emission of thousands of MSPs or emission from cosmic rays injected at the Galactic Centre possible sources \citep{gor14,cal15}. In the MSP scenario, Galactic GCs that are inspiralling towards the Galactic Centre due to dynamical friction are expected to deposit their population of MSPs which are then inherited by the Nuclear Star Cluster and contribute to the observed gamma-ray excess \citep{bra15,abb18,arc18,fao18}. Similarly, a gamma-ray excess was also observed in Andromeda galaxy \citep{fragangn18}.

In this paper, we reconsider the origin and evolution of NSs in GCs. We study the origin and dynamical evolution of the NS population of GCs by the means of direct $N$-body simulations performed with the state-of-the-art collisional evolution code \nbseven\ \citep{aseth2012}. We consider different cluster masses, initial fractions of primordial binaries and metallicities \citep{ban17,ban18}. Finally, we infer the typical population of MSPs in our models and discuss its implications for the MSP origin of the gamma-ray emission in the Galactic Centre.

Our paper is organized as follows. In Section \ref{sect:models}, we describe our numerical models of star clusters. In Section \ref{sect:ns}, we discuss our results and show how the NS population is related to the stellar mass content of the cluster. In Section \ref{sect:msp}, we discuss the gamma-ray emission from MSPs in our clusters. In Section \ref{sect:conc}, we summarize our conclusions.

\section{Models}
\label{sect:models}

\begin{figure*}
  \centering
  \includegraphics[width=.7\linewidth]{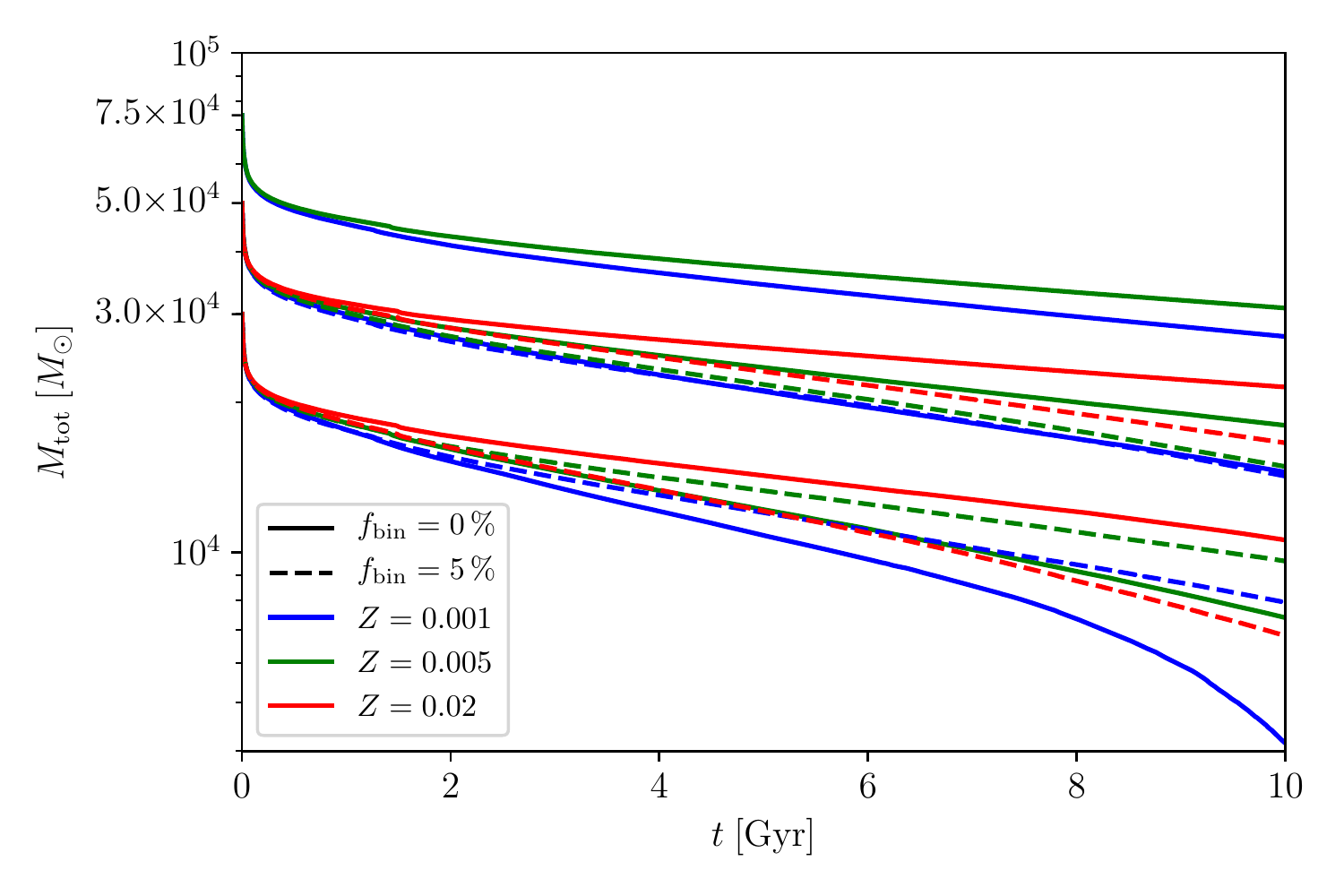}
  \caption{The evolution of the total mass bound to the cluster for all the models in Tab.~\ref{tab:models}. The models may be identified based on their binary fractions (solid lines for $\fbin=0$ and dashed lines for $\fbin=0.05$) and initial masses (specified at $t=0$ Gyr on the vertical axis). Different colours represent different initial metallicities $Z$. From the bottom to the top: models M30k0b and M30k5b ($3.0\times 10^4\,\msun$), models M50k0b and M50k5b ($5.0\times 10^4\,\msun$), and model M75k0b ($7.5\times 10^4\,\msun$).}
  \label{fig:masstot}
\end{figure*}

In the presented work, we utilize direct $N$-body evolutionary models of star clusters with the total initial masses, $\mini$, ranging from $3.0\times 10^4\,\msun$ to $7.5\times 10^4\,\msun$ and the initial half-mass radius $\rhini \approx 2\,\pc$. Stellar masses, $m$, are sampled from the initial mass function \citep[IMF,][]{kro01}, given by
\begin{equation}
\label{eqn:imf}
  \xi(m)=
  \begin{cases}
  k_1\left(\frac{m}{0.08}\right)^{-1.3} & m_\mathrm{min} \leq m/\msun \leq 0.5 \,,\\
  k_2\left(\frac{0.5}{0.08}\right)^{-1.3}\left(\frac{m}{0.5}\right)^{-2.3} & 0.5 \leq m/\msun \leq 100 \,,
  \end{cases}
\end{equation}
where $k_1$ and $k_2$ are the normalization factors and $m_\mathrm{min}=0.08$. The initial binary fraction is defined as
\begin{equation}
	\fbin \equiv \frac{\Nbin}{\Nbin + \Nstar},
\end{equation}
where $\Nbin$ is the number of binaries and $\Nstar$ is the number of single stars in the cluster.
The overall primordial binary fraction in our models is set to either $\fbin=0.0$ or $\approx 0.05$. However, as elaborated in \citet[see their Sec.~2]{ban18}, the initial binary fraction of the O-type stars, which is taken separately, is $\approx100\,\%$, to be consistent with the observed high binary fraction among the O-type stars in young clusters and associations \citep[see, e.g.,][]{Sana_2011}. As discussed there, the O-type-stellar binaries are taken to follow initially the orbital-period distribution of \citet{Sana_2011} and a uniform mass-ratio distribution (an O-star is paired only with another O-star, as it is typically observed, and the pairing among lower-mass stars is obtained separately; see below). The orbital periods of the non-O-star primordial binaries follow the \citet{Duq_1991} distribution that represents a dynamically-processed binary population \citep{Kroupa_1995a}, and their mass-ratio distribution is also taken to be uniform. 
The initial binary eccentricities, $e$, are drawn from a thermal distribution 
\citep{Spitzer},
$f_e(e)=2e$.
Such a scheme of including primordial binaries provides a reasonable compromise between the economy of computing and consistencies with observations \citep{ban18}.
The model parameters are provided in Tab.~\ref{tab:models}.

These long-term direct $N$-body computations, some of which are presented in \citet{ban17} and \citet{ban18}, are performed with \nbseven\ \citep{aseth2012}.
This code is a descendant of the widely-used \nbsix\ direct $N$-body evolution code
\citep{aseth2003,nitadori2012}. \nbseven\ uses
the Algorithmic Regularization Chain \citep[ARC;][]{Mikkola_1999,Mikkola_2008} instead of
the classic KS-Chain Regularization in \nbsix\ \citep{Mikkola_1993,aseth2003}.
This ensures a reliable and numerically stable treatment of multiple systems (subsystems) of arbitrary mass ratios, especially those involving one or more massive objects (e.g.\ BHs), which continue to form and dismantle dynamically in any dense environment. The implementation of ARC also
allows the on-the-fly post-Newtonian (PN) treatment in subsystems of binary or triple BHs when general-relativistic (GR) effects become important \citep{Mikkola_2008}.

\nbseven\ otherwise utilizes similar numerical strategies as \nbsix, i.e.\ a fourth-order Hermite integrator \citep{aseth2003} to accurately advance the trajectories
of each member that is subjected to the resultant force from the rest of the bodies.
To relax the tedious $\propto N^3$ dependence in computing time, a
neighbour-based scheme is applied for evaluating the force contributions
\citep{nitadori2012} at the shortest time intervals (the ``irregular'' force/steps).
At longer time intervals (the ``regular'' force/steps), all members in the
system are included in the force evaluation. Inexpensive (yet numerous) irregular forces
are computed using parallel processing\footnote{OpenMP -- Open Multi-Processing} in regular single-node workstation CPUs, while the much more expensive regular force evaluations are done on {\tt CUDA}\footnote{Compute-Unified Device Architecture}\!-enabled high-performance
GPUs\footnote{All computations in this work are done on
workstations equipped with quad-core {\tt AMD} processors and {\tt NVIDIA}'s
{\tt Fermi} and {\tt Kepler} series GPUs.}\!\!. Diverging Newtonian gravitational forces
during close hyperbolic passages and in hard binaries are handled by two-body or KS regularization
\citep{aseth2003}. Higher-order and GR subsystems are typically treated with the ARC \citep{aseth2012}. For more details, see Sec.~2.1 of \citet[][and references therein]{ban18}.

\setlength\tabcolsep{5pt} 
\begin{table}
\caption{Parameters of the models used: GC mass ($\mini$), half-mass radius ($\rhini$), primordial binary fraction ($\fbin$) and metallicity ($Z$). All models have initial \citet{plummer} distribution.}
\centering
\begin{tabular}{lccccc}
\hline
Name & $\mini\ (\msun)$	& $\rhini\ (\pc)$	& $\fbin$ & $Z$\\
\hline
M30k0b 	& $3.0\times 10^4$		& $2.0$		& $0.00$ & 0.001, 0.005, 0.02\\
M30k5b 	& $3.0\times 10^4$		& $2.0$		& $0.05$ & 0.001, 0.005, 0.02\\
M50k0b 	& $5.0\times 10^4$		& $2.0$		& $0.00$ & 0.001, 0.005, 0.02\\
M50k5b	& $5.0\times 10^4$		& $2.0$		& $0.05$ & 0.001, 0.005, 0.02\\
M75k0b 	& $7.5\times 10^4$		& $2.0$		& $0.00$ & 0.001, 0.005\\
\hline
\end{tabular}
\label{tab:models}
\end{table}
\setlength\tabcolsep{6pt} 

We consider three different initial cluster masses of $3.0\times 10^4\,\msun$, $5.0\times 10^4\,\msun$ and $7.5\times 10^4\,\msun$. Fig.~\ref{fig:masstot} illustrates the evolution of the total mass bound to the cluster for all the models in Tab.~\ref{tab:models}; Tab.~\ref{tab:10gyr} reports the final masses of all clusters, i.e.\ at 10\,Gyr. The clusters lose roughly two thirds of the their initial mass by that time. The final mass depends on the initial features of each cluster. Different initial fractions of primordial binaries and metallicities translate into a factor of about $2$ to $3$ in the final mass.

\section{Neutron stars formation and evolution}
\label{sect:ns}

\begin{table}
\caption{The number of NSs in our simulations and the total mass of the clusters, both at $10\,\gyr$.}
\centering
\begin{tabular}{lcrc}
\hline
Model & $Z$	& $n_\mathrm{NS}$ & $M_\mathrm{tot} (\times 10^4\,\msun)$ \\
\hline
M30k0b & 0.001 & 28 & 0.41 \\
   	   & 0.005 & 35 & 0.74 \\
   	   & $0.02\,\ $ & 34 & 1.06 \\
M30k5b & 0.001 & 44 & 0.79 \\
   	   & 0.005 & 47 & 0.96 \\
   	   & $0.02\,\ $ & 32 & 0.68 \\
M50k0b & 0.001 & 77 & 1.45 \\
   	   & 0.005 & 75 & 1.80 \\
   	   & $0.02\,\ $ & 54 & 2.14 \\
M50k5b & 0.001 & 74 & 1.42 \\
   	   & 0.005 & 69 & 1.48 \\
   	   & $0.02\,\ $ & 50 & 1.66 \\
M75k0b & 0.001 & 110 & 2.70 \\
   	   & 0.005 & 128 & 3.08 \\
\hline
\end{tabular}
\label{tab:10gyr}
\end{table}

\begin{figure*}
    \centering
    \hfill
    \includegraphics[width=.46\linewidth]{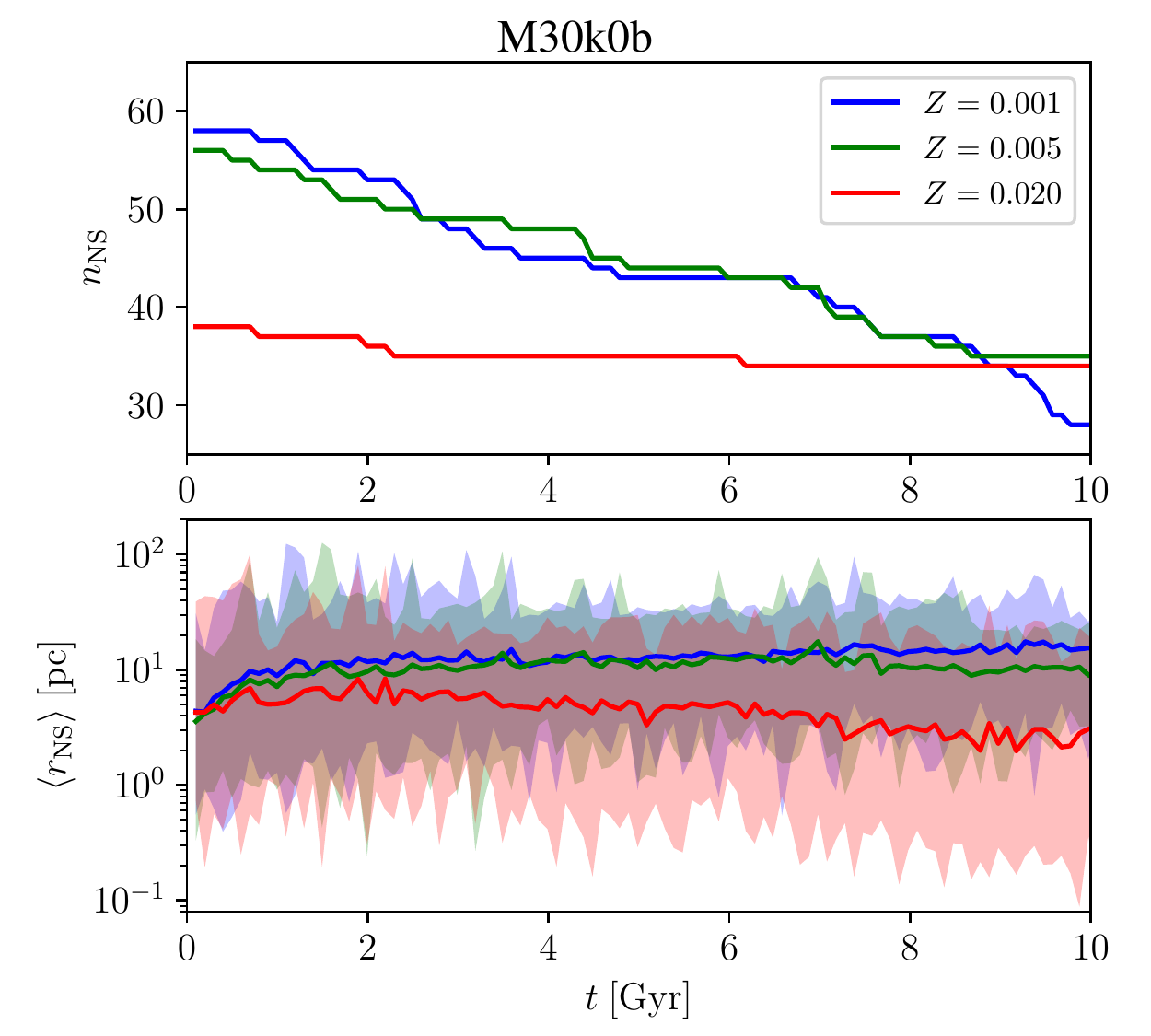}
    \hfill
    \includegraphics[width=.46\linewidth]{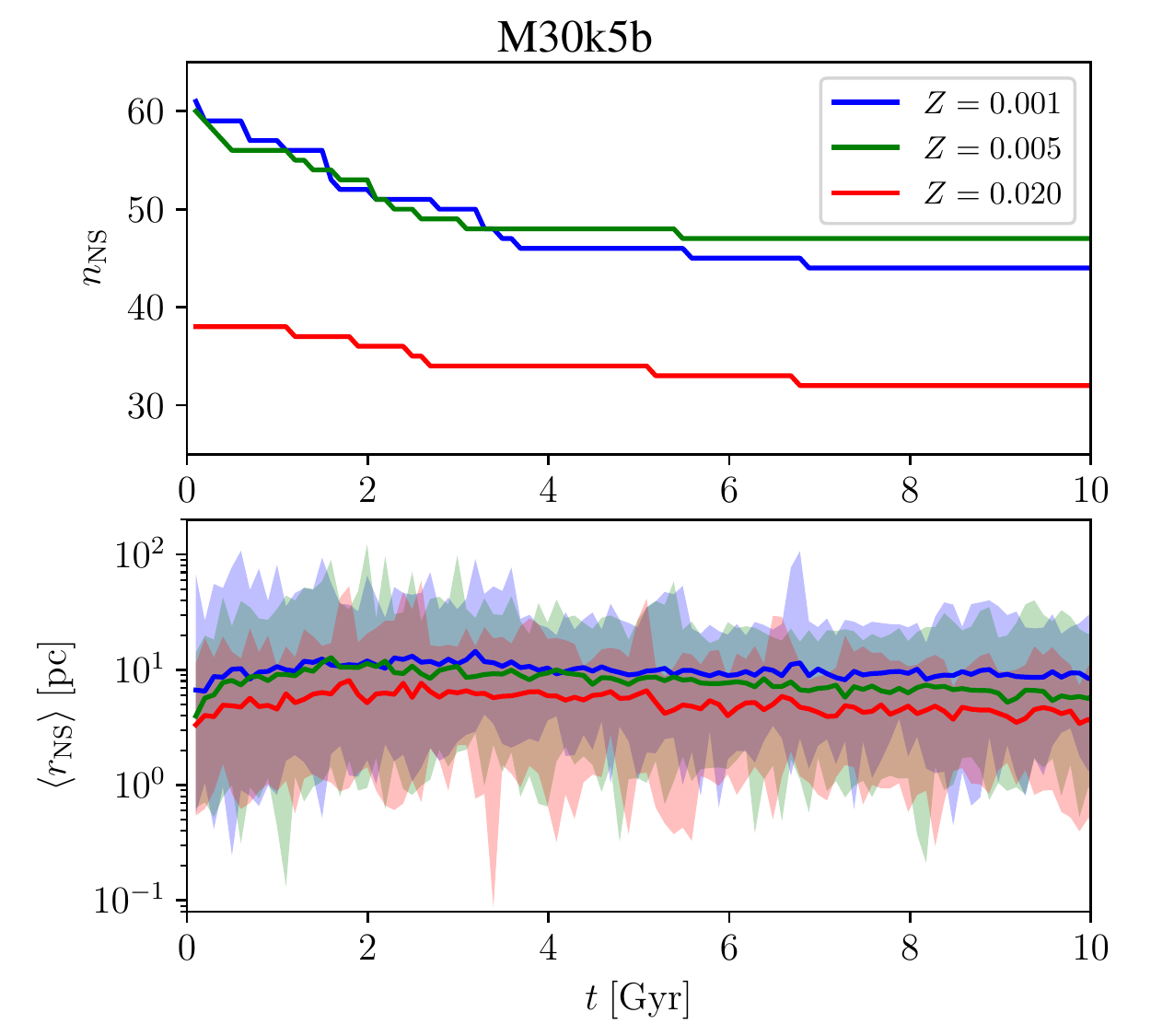}
    \hfill\\
    \hfill
    \includegraphics[width=.46\linewidth]{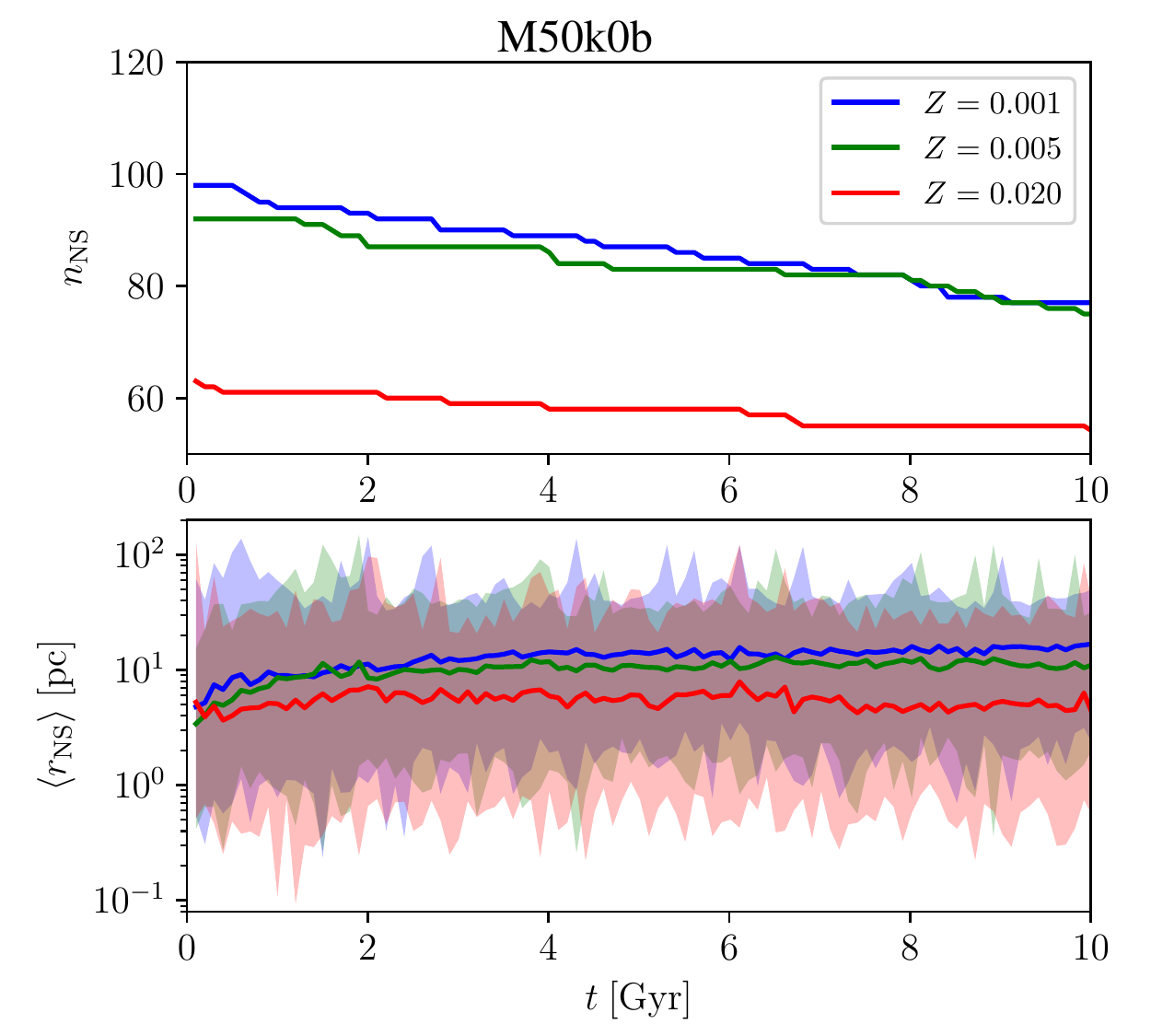}
    \hfill
    \includegraphics[width=.46\linewidth]{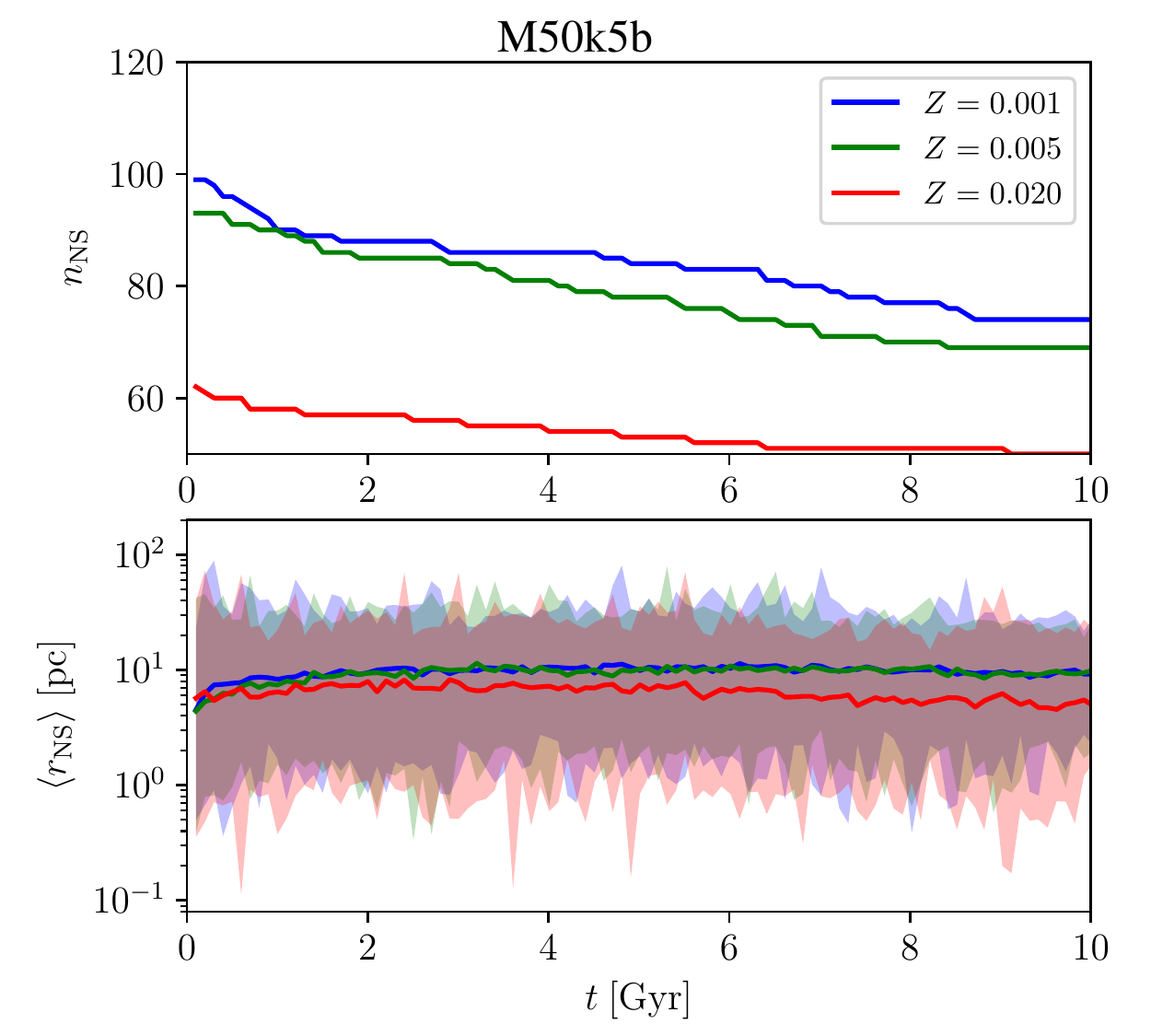}
    \hfill\\
    \includegraphics[width=.46\linewidth]{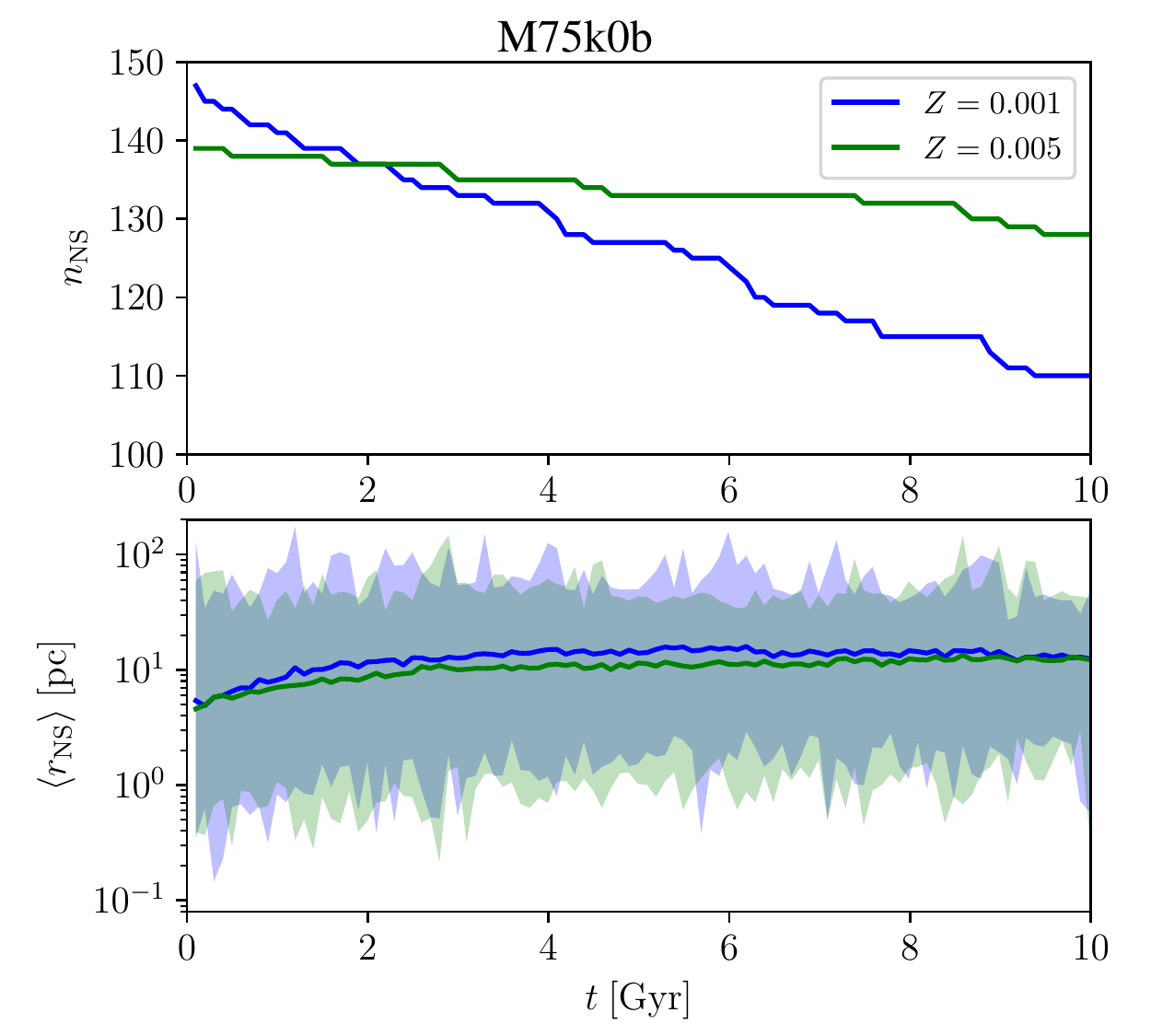}
    \caption{Time evolution of the number of NSs and the average distance of NSs from the GC centre for models M30k0b (top-left), M30k5b (top-right), M50k0b (middle-left), M50k5b (middle-right) and M75k0b (bottom-centre). Upper panels: The total number of NSs bound to the cluster in our models (see Tab.~\ref{tab:models}) for different values of the stellar metallicity. Lower panels: Solid lines represent the mean distance from the GC centre of the bound NSs; each highlighted area is the distance range of all the NSs.}
	\label{fig:ns}
\end{figure*}

\begin{figure*}
  \centering
  \includegraphics[scale=1]{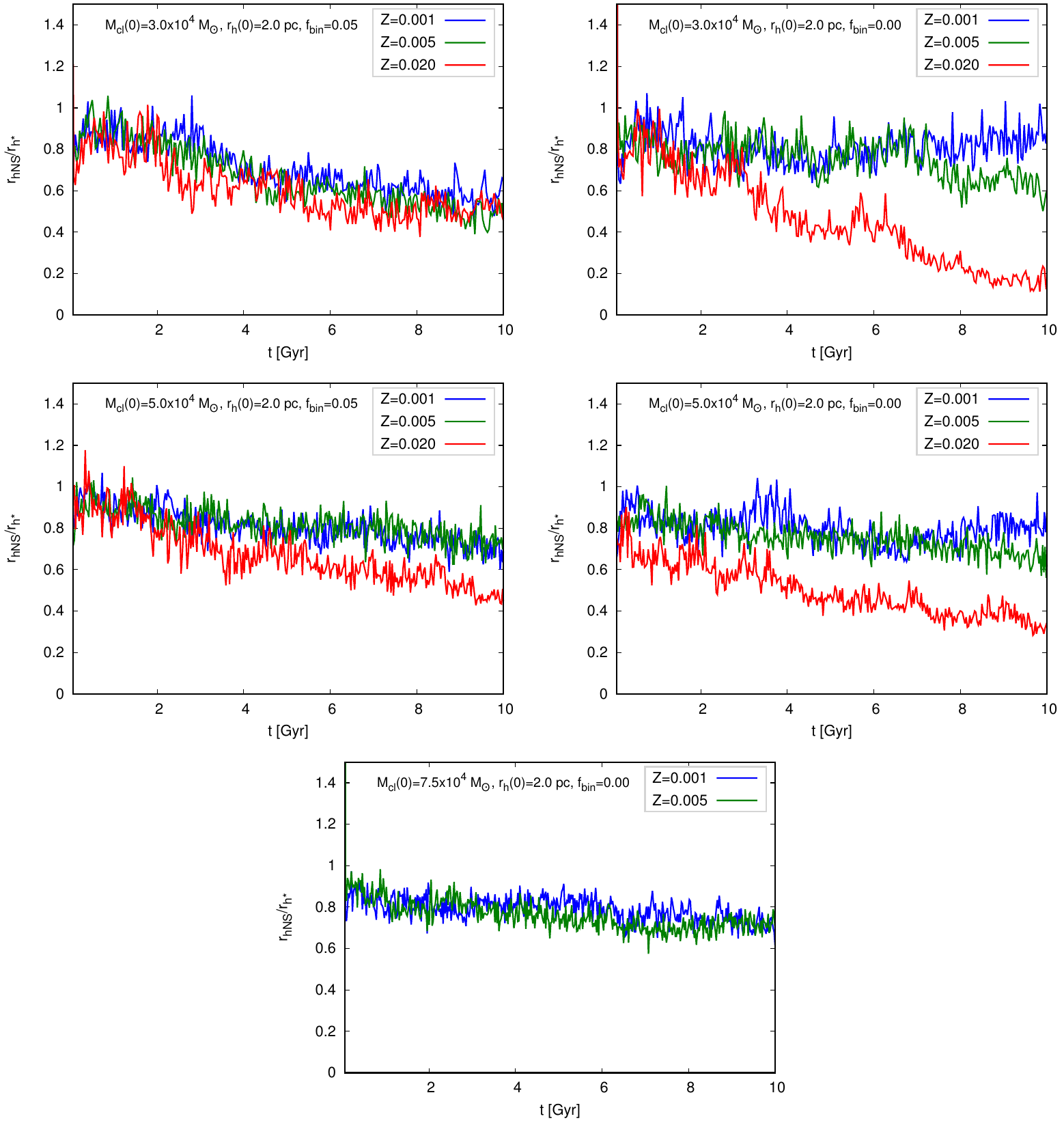}
  \caption{The ratio of the average distance of NSs from the GC centre and the half-mass radius for models M30k0b (top-left), M30k5b (top-right), M50k0b (middle-left), M50k5b (middle-right) and M75k0b (bottom-centre).}
  \label{fig:rnsrhratio}
\end{figure*}

In the models presented here, we used the single-star and binary-star evolution models (\sse, \bse) of \citet{hur00,hur02} that are integrated in the \nbseven\ code. That way, dynamical evolution and interactions are computed in tandem with the nuclear evolution of individual members. The \sse\ algorithm is parametrized by the body's mass $m$ and metallicity $Z$, and it is able to describe the evolution in a mass range $m \in (0.1,100)\,\msun$, by default. In the computed models, however, the strict $100\,\msun$ limit is relaxed by adopting an optimally-sampled IMF \citep{Kroupa_2013}. Generally, the nuclear evolution is faster for stars of higher-mass and lower metallicity. The exact time until a main sequence star becomes a NS, however, also depends on whether it is in a binary or not because the \bse\ introduces additional physical processes, e.g.\ mass transfer, accretion, collisions etc. \citep[cf.][]{1992ApJ...391..246P,2003NewA....8..817D,zapartas2017,pav18} 

The process of the final transformation from a giant star to a NS is usually accompanied by a type-II supernova (SN) explosion. As the supernova explosion is never perfectly symmetric, there is an excess of momentum of the ejecta in one direction which gives a recoil velocity to the remnant, i.e.\ \emph{a velocity kick} \citep{lyn94}. The distribution of kicks is supposed to be Maxwellian with a dispersion of $\sigma \approx 190\,\kms$ \citep{han97} or $\sigma \approx 300\,\kms$ as in \citet{Hobbs_2005} observations.
Based on the mass of a star before becoming a SN, the explosion can either form a NS or a BH. The latter could also receive a kick, however, its properties may be different \citep[cf.][]{bel08,ver17,pav18}. Another way of forming a NS is via an electron-capture supernova \citep[ECS;][]{Podsiadlowski_2004}. The ECS-NSs would receive no kick at birth or only a very small one due to an asymmetry in neutrino emissions. Recent hydrodynamic studies of the ECS-NS formation also support very small natal kicks \citep{Gessner_2018}.

The production of NSs and the evolution of their population in our models is presented in Fig.~\ref{fig:ns}. In the top panels, we plot the total number of NSs in each cluster in time. In the low-metallicity realisations, remnants are produced at earlier times because of a lower mass limit for the stellar core collapse. Therefore, we observe a larger number of NSs at the beginning of those calculations.

The NSs which form during a type-II SN mostly leave the cluster shortly after their birth because the escape velocity from the modelled clusters is much lower than the kick velocity that they receive, because it is drawn from a distribution with a rather high dispersion $\sigma$. The NSs that remain bound to the cluster are therefore mainly produced via ECS.
Note that only in an environment comparable to a Nuclear Cluster, where the escape velocity is of the same order as $\sigma$, the high-kick NSs could remain bound \citep[see e.g.][]{ban17,ban18}.

\begin{figure*}
  \centering
  \includegraphics[width=.7\linewidth]{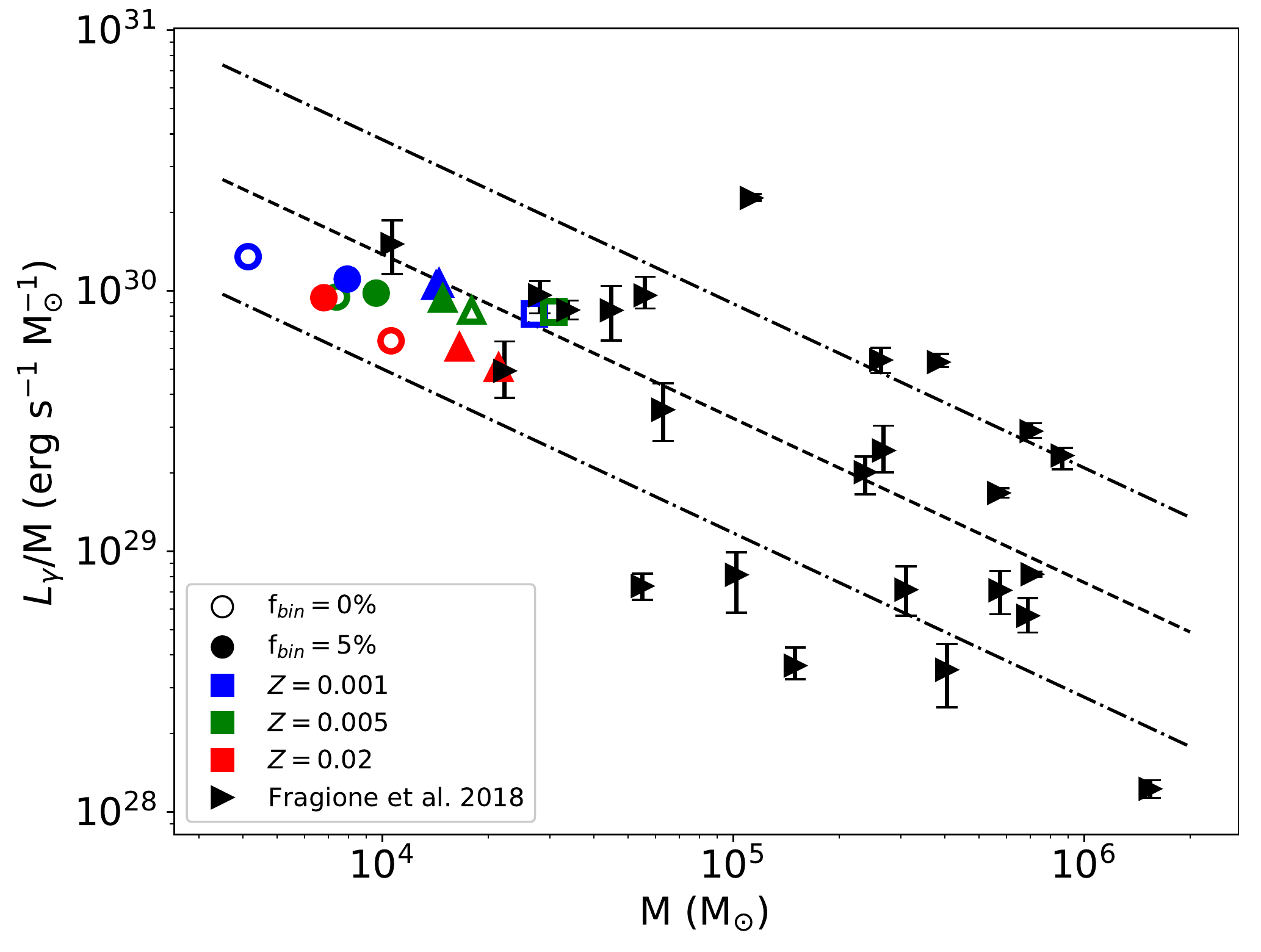}
  \caption{Ratio of gamma-ray luminosity to globular cluster mass $L_\gamma/M_\mathrm{GC}$ as a function of the cluster mass. The coloured symbols (circles: $\mini=3.0\times 10^4\msun$; triangles: $\mini=5.0\times 10^4\msun$; squares: $\mini=7.5\times 10^4\msun$) represent the inferred $L_\gamma/M_\mathrm{GC}$ from the different GCs studied in this work, by assuming a recycling fraction $f_\mathrm{rec}=0.1$ \citep{abb18} and an average MSP gamma-ray emission of $L_\gamma=2\times 10^{33}$\,erg\,s$^{-1}$ \citep{bra15}. The dashed line shows the best log-linear fit to the data \citep*[as in][]{fao18} and the dashdotted lines show $1\sigma$ deviations.}
\label{fig:lmcl}
\end{figure*}

Although the bound NSs are, on average, heavier and remain more concentrated than regular stars, their mass segregation is also partly quenched due to a heating effect of the BHs. Initially, i.e.\ when the BH-engine is active, we can see an overall expansion of the NS population\footnote{Stellar population expands as well.} (bottom panels of Fig.~\ref{fig:ns}). They begin to collapse back when the engine is weakened, i.e.\ when a majority of BHs is dynamically ejected. For a lower $Z$, the BHs may become more massive and more numerous, resulting in higher energy deposition in the rest of the cluster members. Consequently, the expansion is faster and the re-collapse occurs later.

Fig.~\ref{fig:rnsrhratio} illustrates the ratio of the average NSs distance from the GC centre and the half-mass radius, i.e.\ $r_\mathrm{hNS}/r_\mathrm{h^*}$. As it has been discussed, the radial profile of the NSs depends on the BH cluster activity and dynamical ejection, which in turn depends on the metallicity. However, we find that the NS population always approximately ``maps'' the stellar population, in the sense that the ratio $r_\mathrm{hNS}/r_\mathrm{h^*}$ decreases from nearly unity to $\approx 0.6-0.8$ in a few Gyrs in all the clusters, independently of the initial metallicity, binary fraction, and cluster mass. Exceptions to the overall behaviour are the solar-metallicity models with zero primordial binaries. In these models, the segregation of the NSs is stronger due to a weaker BH heating and also the lack of heating from the primordial binaries. However, $\fbin=0$ is probably an idealization since all clusters would have been born with a non-negligible primordial-binary population \citep{mak12}, which partially suppresses the NS segregation even when the BH heating is weak.

\section{Millisecond pulsars, gamma-ray emission and the Galactic Centre gamma-ray excess}
\label{sect:msp}

Recycling a NS in order to produce a MSP requires a small orbital separation in the progenitor binary. Dense stellar environments may help forming tight binaries, either via direct collision with a giant, tidal capture or exchange encounters \citep[see, e.g.,][]{Hut_1992,Banerjee_2007}. \citet{fab75,fab83} first predicted that MSPs would be abundant in GCs due to a high rate of stellar dynamical encounters
\begin{equation}
\Gamma \propto \frac{\rho_\mathrm{c}^2 r_\mathrm{c}^3}{\sigma_\mathrm{disp}} \,,
\end{equation}
where $\rho_\mathrm{c}^2$, $r_\mathrm{c}^3$ and $\sigma_\mathrm{disp}$ are the core density, the core radius and the velocity dispersion, respectively. Recent measurements by \citet{bah13} confirmed that there is a correlation between the number of X-ray and radio sources in a given GC and its stellar dynamical encounter rate $\Gamma$, while \citet{ver14} showed instead that the typical encounter rate for a single binary plays a role in providing a good characterization of the differences between the pulsar populations in various GCs.

Despite the success in explaining the observed features of X-ray binaries and MSPs, a lot of uncertainties still remain, e.g.\ the cluster's IMF, the primordial binary population, the typical lifetime of binary progenitors of MSPs and, in particular, the retention fraction of NSs \citep{pod04,iva08,ver14}. Moreover, the physics of LMXBs is usually modelled with simplified assumptions, which only recently have been discussed in more detail by \citet{tau11} and \citet{tau12}.

Recently, new attention to MSPs has been brought up by \textit{Fermi-LAT} telescope\footnote{https://fermi.gsfc.nasa.gov/}\!\!, which provided high-quality data in the energy range from $20$\,MeV to over $300$\,GeV. The data show a gamma-ray excess around the Galactic Centre. Its peak energy is at $\approx 2$\,GeV, with an approximately spherical density profile decreasing as $r^{-2.4}$ out to $3$\,kpc from the Galactic Centre \citep{aba14,cal15}. There is no clear understanding of the origin of such a gamma-ray excess, with dark matter annihilation and MSPs possible sources. While the first interpretation is challenged by a non-detection of the corresponding signal from dwarf spheroidal satellite galaxies of the Milky Way \citep{alb17}, the emission of thousands of unresolved MSPs seems to be the most promising explanation \citep{lee15,bar16,ack17}. \citet{bra15} proposed that the population of MSPs was left in the inner regions of the Milky Way as a consequence of GCs migration and disruption due to dynamical friction.

\citet*{fao18} used the data from \citet{hoo16} to derive a relation between the gamma-ray emission from GCs and their masses. They found that the gamma-ray emission from GCs, as measured by \textit{Fermi}, is well described by the following log-linear relation 
\begin{equation}
  \log(L_\gamma/M_\mathrm{GC})=32.66\pm 0.06 -(0.63\pm 0.11) \log(M_\mathrm{GC}) \,,
  \label{eqn:lgmc}
\end{equation}
where $L_\gamma$ is the gamma-ray emission of a GC and $M_\mathrm{GC}$ is its mass. Fig.~\ref{fig:lmcl} shows the ratio of the gamma-ray luminosity to the globular cluster mass, $L_\gamma/M_\mathrm{GC}$, as a function of the cluster mass \citep[as in][]{fao18} and the inferred $L_\gamma/M_\mathrm{GC}$ of the GCs studied in this work. In deriving $L_\gamma/M_\mathrm{GC}$ for our models, we use the final mass of our evolved GCs and the final number of NSs as was found in our simulations (see Tab. \ref{tab:10gyr}). We also assume a recycling fraction, i.e.\ the number of NSs that have been spun up, of $f_\mathrm{rec}=0.1$ as found by \citet{abb18} and an average MSP gamma-ray emission of $L_\gamma=2\times 10^{33}$\,erg\,s$^{-1}$ \citep{bra15}. We find that our GCs models are in a good agreement with Eq.~\eqref{eqn:lgmc}. By using this equation and evolving the primordial GC population in the Milky Way potential \citep*[see also][]{fgk18} and taking into account the MSP spin-down, \citet{fao18} showed that the accumulated population of MSPs well reproduces both in intensity and spatial profile the gamma-ray excess observed by \textit{Fermi}. Since our inferred $L_\gamma/M_\mathrm{GC}$ are in a good agreement with Eq.~\eqref{eqn:lgmc}, we conclude that our results support the MSP origin of the gamma-ray excess signal observed by \textit{Fermi} in the Galactic Centre.

\section{Discussions and Conclusions}
\label{sect:conc}

NSs are among the most interesting astrophysical objects, being precursors of a lot of high-energy phenomena, such as gravitational waves and gamma-ray bursts. GCs are the natural environment where hundreds of NSs can form and dynamically evolve. As a consequence of high stellar densities and low velocity dispersions, NSs can interact with single and binary stars in GCs more frequently than in any other environment, and form bound binary systems, where they are observed either as LMXBs or MSPs in X- and gamma-rays \citep{iva08,kat75,ver14}.

We studied the origin and dynamical evolution of NSs within GCs with different initial masses, metallicities and binary fractions. We found that even though hundreds of NSs are formed in a star cluster, most of them receive a velocity kick at birth such that they escape from their parent GC \citep{dav98,tre10}. The radial profile of NSs is shaped by the BH content of the cluster, which partially quenches the NS segregation, potentially connected to gravitational waves generation \citep{ban17,ban18,frak18}. We also found that independently on the cluster mass and initial configuration, the NSs reasonably map the average stellar population, as their average radial distance is $\approx 60-80$\% of the cluster half-mass radius.

MSPs are believed to be recycled pulsars, where a NS spins up by accreting mass from a companion star in a binary system \citep{phi94,tau12}. New attention to MSPs comes from the recent high-quality data by \textit{Fermi}, which observed an extended gamma-ray emission around the Galactic Centre. One of the possible origin of the excess is the emission by thousands of MSPs, probably delivered by Galactic GCs that inspiralled onto the Galactic Centre due to dynamical friction \citep{bra15,fao18}. By assuming a recycling fraction $f_\mathrm{rec}=0.1$ \citep{abb18} and an average MSP gamma-ray emission of $L_\gamma=2\times 10^{33}$\,erg\,s$^{-1}$ \citep{bra15}, we computed the $L_\gamma/M_\mathrm{GC}$ of the GCs studied in this work and found that our results support the MSP origin of the gamma-ray excess signal observed by \textit{Fermi} in the Galactic Centre.

We note that the inferred $L_\gamma$ for a given cluster primarily depends on the number of NSs that are retained for a long term in the cluster, i.e.\ the number of ECS NSs, and also on the recycling fraction and the typical gamma-ray luminosity per NS. The latter two ingredients may change from cluster to cluster and are highly uncertain observationally (e.g.\ the shape of the luminosity function and its boundaries), and typically available only for a few Galactic clusters. Other various initial ingredients, such as the IMF and binary properties, may affect the resulting $L_\gamma/M_\mathrm{GC}$. In our calculations, we took all of them in agreement with observations (see Sect.~\ref{sect:models} and Sect.~\ref{sect:msp}). Among these, assuming small natal kicks of ECS NSs, instead of zero as in the present work, we would have ejected some of the ECS NSs and $L_\gamma$ would have been smaller. On the other hand, a top-heavy IMF, although it would produce a larger number of ECS NSs, would also have aided the host cluster's dissolution (due to both larger stellar-evolutionary mass loss and more important BH heating). Consequently, $L_\gamma/M_\mathrm{GC}$ would have been larger over the cluster's lifetime. Finally, we note that we considered initial cluster masses up to $7.5\times 10^4 \msun$. Direct $N$-body simulations of large clusters are still almost impossible at the present day, in particular if primordial binaries are present (as in this study) and are beyond the scope of this work. Future studies shall be aimed both at simulating larger clusters and at understanding the exact role of all the relevant assumptions.

\section{Acknowledgements}

This research was partially supported by an ISF and an iCore grant. GF acknowledges support from an Arskin postdoctoral fellowship and Lady Davis Fellowship Trust at the Hebrew University of Jerusalem. VP acknowledges support from Charles University, grants GAUK-186216 and SVV-260441. This work has been partly supported by the Deutsche Forschungsgemeinschaft (DFG; German Research Foundation) through the individual research grant "The dynamics of stellar-mass black holes in dense stellar systems and their role in gravitational-wave generation" (BA 4281/6-1).
SB is thankful to the computing team of the Argelander-Institut f\"ur Astronomie, University of Bonn, for their efficient maintenance of the workstations on which all the computations have been performed.

\bibliographystyle{mn2e}
\bibliography{refs}
\end{document}